\documentclass[amsmath, amssymb, aps, prb, twocolumn, superscriptaddress, longbibliography, hypertext]{revtex4-2}
\usepackage[utf8]{inputenc}
\usepackage[english]{babel}
\usepackage[T1]{fontenc}
\usepackage{graphicx}
\usepackage[usenames]{color}
\usepackage{hyperref}
\makeatletter
\newcommand*{\rom}[1]{\expandafter\@slowromancap\romannumeral #1@}
\makeatother
\newcommand{\Figref}[1]{Fig.~\ref{#1}}
\newcommand{\Eqref}[1]{Eq.~(\ref{#1})}

\newcommand{\dipc}{Donostia International Physics Center (DIPC), E-20018 Donostia-San Sebastián, Spain}

\begin{document}

\title{Non-Hermitian Effects in the Su-Schrieffer-Heeger model: Exploring Substrate Coupling and Decoupling Dynamics}

\author{Shayan Edalatmanesh}
\affiliation{\dipc}

\author{Thomas Frederiksen}
\affiliation{\dipc}
\affiliation{Ikerbasque, Basque Foundation for Science, E-48013 Bilbao, Spain}
\begin{abstract}
The substrate-adsorbate interaction can significantly influence the adsorbate’s electronic structure, stability, reactivity, and topological properties. In this study, we investigate the emergence of non-Hermitian physics in the Su--Schrieffer--Heeger (SSH) model when coupled to a substrate, focusing on the impact of substrate interaction on the electronic states of the adsorbate. We demonstrate how the coupling between the SSH chain and the underlying substrate induces non-Hermitian effects, which manifest as amplification or attenuation of zero-energy electronic states. Furthermore, inspired by novel experimental techniques such as using a scanning tunneling microscope tip to lift part of the nanomaterial, we present simulations of scenarios where a segment of the SSH chain is decoupled from the substrate. By examining various configurations, including cases with odd or even numbers of sites coupled to the substrate, we demonstrate that tuning the coupling strength induces novel phenomena, such as the emergence of a zero-energy monomode or additional zero-energy states localized at the boundary between on-surface and suspended chain segments. Our results reveal the role of substrate coupling in shaping the topological properties of non-Hermitian SSH chains, offering new insights into tunable non-Hermitian effects and their potential applications in quantum technologies and nanodevices.
\end{abstract}

\maketitle{}

\section{Introduction}

An effective implementation of quantum materials in cutting-edge technological devices has the potential to dramatically improve our present-day methods of data processing, storage and transfer. Thus, the development of new materials with tailored and enhanced electronic, magnetic, and optical properties is of paramount importance. In this regard, topological quantum materials (TQMs) are attractive candidates for the next generation of quantum devices. Topological insulators can exhibit various exotic quantum phenomena such as quantum anomalous Hall effect \cite{Sci-doi:10.1126/science.1234414}, Majorana fermions \cite{PhysRevLett.111.147202} -- potentially useful as quantum building blocks (qubits) in fault-tolerant quantum computing \cite{Weng_2015,Sarma2015} -- and axion insulators \cite{Liu2020}. Their robust topological phases, low bandgap and symmetry-protected, zero-energy electronic states give them stability against decoherence and hold promise of impacting various fields such as electronics and spintronics. 

Modern TQMs can be experimentally synthesized and studied using on-surface synthesis (OSS) \cite{Grill2007, Grill2020} in combination with scanning probe microscopy techniques \cite{giessibl2003, ALLERS1999247, Binnig1982, dimas2013, Pavlicek2017, Sugimoto2007, Stetsovych2017}. So far, in this relatively new subdiscipline of surface science, most attempts have been limited to the synthesis and study of physisorbed, planar TQMs (i.e., polymers with $sp^2$ hybridization and fully saturated bonds) \cite{Chen2013, Ruffieux2016,Rizzo2018,Cirera2020,gonzalez21, Cai2010, song2022designer, Li2021, Sun2021}, treating the substrate as a bed that drives the chemical reactions yet does not affect the topological properties of the material. Nonetheless, real-world application of TQMs requires their integration into devices and interfaces, which necessitates the investigation of the environment-induced effects that could potentially influence their electronic structure and topological properties.
 
The simplest framework to study a topological system is the Su--Schrieffer--Heeger (SSH) model \cite{ssh.PhysRevLett.42.1698}. The SSH model, originally introduced to describe the electrical conductivity increase in doped polyacetylene \cite{ssh}, describes a half-filled one-dimensional ($1$D) lattice with two sites in every unit cell. In this model, electrons are allowed to hop between sites within the same unit cell (intra-cell hopping) as well as between neighboring sites in adjacent unit cells (inter-cell hopping). Depending on the ratio of the two different hopping energies, the SSH chain can either be classified as a topologically trivial or non-trivial phase. The key distinction is that the non-trivial phase, under open boundary conditions, hosts robust zero-energy edge states that are topologically protected by chiral symmetry. \Figref{fig:fig1}(a, b) illustrates schematics corresponding to the two topological phases, highlighting their distinguishing features.

In its conventional form, the SSH Hamiltonian is Hermitian, ensuring real eigenvalues and a well-defined topological invariant, see \Figref{fig:fig1}(c, d). However, in a realistic device setting on a metal surface, the system cannot be completely isolated from its environment. Connecting the SSH chain (as an adsorbate) to a substrate (or a lead) induces a self-energy on the system, and in an effective picture we can write the Hamiltonian as
\begin{equation}\label{eqn:hamil}
H_{\text{eff}} = H_{\text{SSH}} + \Sigma(E),
\end{equation}
where $H_{\text{SSH}}$ is the Hamiltonian of the isolated system and $\Sigma(E)$ reflects the coupling to the environment (e.g., a substrate, a tip probe, or a combination of reservoirs) via self-energies at energy $E$ close to the Fermi level. Since $\Sigma$ is of complex nature (with an imaginary on-site energy contribution), $H_{\text{eff}}$ is now non-Hermitian (NH) \cite{Moiseyev2011}. 

Unlike Hermitian systems, which are characterized by real eigenvalues and conventional topological phases, NH Hamiltonians can possess complex energy eigenvalues, exceptional points (EPs), and unconventional topological behavior. In the physical sense, the presence of complex eigenvalues ($\epsilon = E \pm i\gamma$) reflects the existence of quasi-bound states in the open system that can either gain or lose energy at a rate determined by $\gamma (\neq 0)$, linking the system to its environment (i.e., a reservoir) \cite{Avila2019, ernzerhof}. EPs, in turn, are singularities in the parameter space where two or more eigenstates - both their eigenvalues and eigenvectors - coalesce, giving rise to fundamentally altered physical properties not captured by conventional Hermitian frameworks.

In particular, EPs can potentially be harnessed to engineer robust modes, unusual transport phenomena, and enhanced sensing capabilities. They could influence the nature of topological phases in NH lattices, impact bulk-boundary correspondence, and enable the design of non-reciprocal devices and exotic localization phenomena that have no direct analogs in Hermitian settings \cite{Zhen2015,Bergholtz2021,Budich2019,Okugawa2019}. Such EP physics thus provides a versatile platform for exploring new states of matter and novel device functionalities in NH condensed matter systems.

In this study, we investigate how coupling to a substrate can serve as a source of non-Hermiticity in SSH chains. We consider both weak and strong coupling between an on-surface SSH chain and a substrate, as well as partial decoupling of the adsorbate (lift-off with an STM tip). We show how the different coupling regimes affect electronic states, their localization in the real space, and the emergence of EPs along with other topological features in the SSH model. Our work uncovers novel NH boundary effects in SSH chains and offers insights into experimentally tuning and realizing such systems.

\section{SSH chains with and without coupling to a substrate}

All the numerical calculations of non-Hermitian tight-binding models in this work have been performed using the sisl library \cite{zerothi_sisl}, a Python-based package for atomistic simulations. The Hamiltonians were diagonalized using the SciPy package \cite{sciPy-NMeth}, and the resulting data was visualized with Matplotlib \cite{Hunter:2007}.

\subsection{Hermitian SSH Model}
\begin{figure}
\centering
\includegraphics[width=\columnwidth]{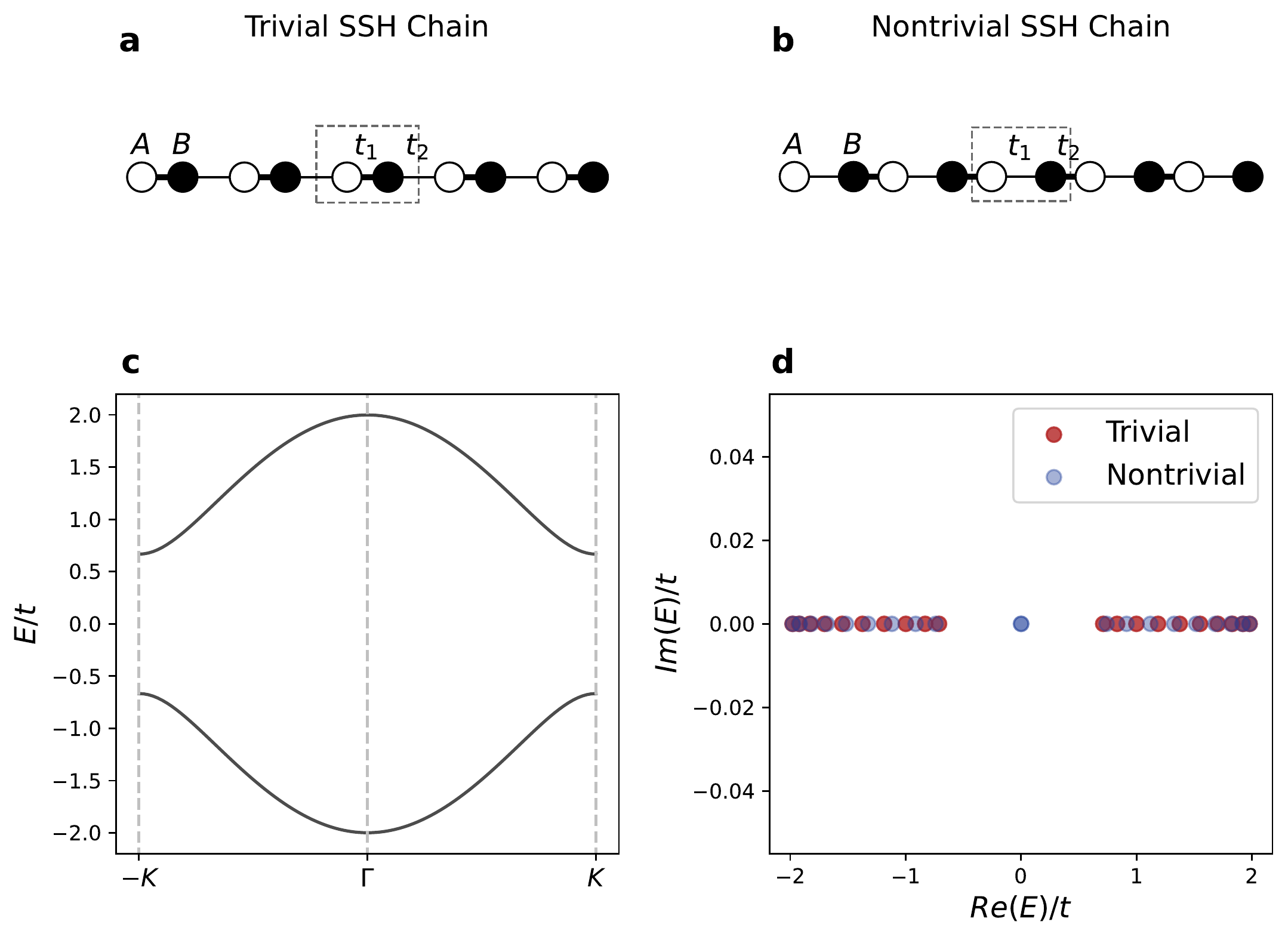}
\caption{Schematic representations and corresponding energy spectra of the SSH model in its trivial and nontrivial phases. (a) The trivial configuration, characterized by an alternating hopping pattern with intercell hopping amplitude $ |t_1| > |t_2| $, and (b) the topologically nontrivial configuration, with $ |t_1| < |t_2| $. (c) The bulk band structure $ E(k) $ for the trivial ($t_1=4/3t$ and $t_2=2/3t$) and nontrivial ($t_1=2/3t$ and $t_2=4/3t$) phases are identical, with the presence of a band gap. (d) The eigenvalue spectra in the complex energy plane of the finite chain of $N=10$ unit cells for both classes of topologically trivial and nontrivial. The energies remain real for both phases, confirming the Hermitian nature of the system.}
\label{fig:fig1}
\end{figure}

The standard SSH model is defined as a chain composed of $N$ unit cells, each containing two sublattice sites, $A$ and $B$. The Hamiltonian in real space can be written as:
\begin{equation}
H_{\text{SSH}} = \sum_{n}^{N}\bigl( t_1 |n,A\rangle\langle n,B| + t_2 |n,B\rangle\langle n+1,A| + \text{h.c.}\bigr),
\end{equation}
where the index $n=1,\dots,N$ labels the unit cells, $t_1$ and $t_2$ are the intracell and intercell hopping amplitudes, respectively, and ``h.c.'' denotes Hermitian conjugation. The aforementioned hopping parameters can also be defined with respect to an average hopping energy $t$, incorporating a modulation term $\delta{t}$. This is expressed as
\begin{equation}
    t_i = t \pm \delta{t},
\end{equation}
where $i = {1, 2}$. In the following we fix $\delta t = \pm t/3$ and thus measure all energies in units of $t$.

For large $N$ or under periodic boundary conditions, it is convenient to move the calculations to momentum space. 
In this basis, the Hermitian SSH Hamiltonian is
\begin{equation}
H_{\text{SSH}}(k) = \begin{pmatrix}
0 & t_1 + t_2 e^{-ik} \\
t_1 + t_2 e^{ik} & 0
\end{pmatrix},
\end{equation}
with the eigenvalues
\begin{equation}
E_{\pm}(k) = \pm \sqrt{t_{1}^{2} + t_{2}^{2} + 2 t_{1} t_{2} \cos(k)}.
\end{equation}

The periodic band structure for this model is illustrated in \Figref{fig:fig1}(c),
where it is seen that the electronic band gap is given by $4 \delta{t}$ and the total band width by $4 t$. As shown in \Figref{fig:fig1}(d), which represents the SSH chain with $N=10$ unit cells under open boundary condition, the energy eigenvalues for both topological classes remain entirely real in the case of the Hermitian SSH model.

\subsection{Non-Hermitian Extension via Self-Energies}

It is noteworthy that the conventional notions of topological triviality or non-triviality, as established in Hermitian systems, rely on real-valued spectra and symmetries preserved by Hermitian Hamiltonians. In our non-Hermitian case, the presence of gain and loss invalidates these assumptions, rendering the hermitian-defined, standard topological invariants (e.g., winding number, Zak phase or Chern number) inapplicable. 

Although NH extensions of topological theory do exist, they involve specific tools and definitions, such as biorthogonal polarization \cite{Kunst2018,edvardson2020, Bergholtz2021}, complex Berry phase \cite{lieu2018}, generalized Brillouin Zone \cite{songfeyyao}, non-Bloch wavefunctions \cite{Yokomizo2019} or modified Chern numbers \cite{yaoChern2018}. Since our focus here is not on developing these advanced NH topological classifications, we will not employ the terminology of topologically “trivial” or “non-trivial” phases for our NH-SSH model. Instead, we acknowledge that the NH regime calls for a distinct topological framework, and we refer the interested reader to provided references for a comprehensive discussion.

To systematically explore the role of non-Hermiticity in our $1$D SSH chains on surfaces, 
we define two distinct regimes for the hopping amplitudes:
\begin{itemize}
    \item \textbf{Case~1:} $|t_1| > |t_2|$  (``trivial'' in the Hermitian limit)
    \item \textbf{Case~2:} $|t_2| > |t_1|$  (``nontrivial'' in the Hermitian limit)
\end{itemize}
Throughout the remainder of this work, we will refer to these as Case~1 and Case~2, respectively, to emphasize that our focus extends to NH extensions of the SSH model.

To incorporate the influence of the substrate we introduce a self-energy term $\Sigma(E)$ to the on-site terms of the SSH Hamiltonian. Formally, the Green's function of the chain is
\begin{equation}
G(E) = [E - H_{\text{SSH}} - \Sigma(E)]^{-1},
\label{eqn:gFunction}
\end{equation}
where the self‐energies mimic the broadening and shift of adsorbate states due to coupling with a continuum of substrate states in the spirit of Newns--Anderson model \cite{An.1961.LocalizedMagneticStates, Ne.1969.Self-ConsistentModelof}
\begin{equation}
\Sigma (E) = -i \Gamma_\text{sub} (E)/2.
\label{eqn:sigma}
\end{equation}

\begin{figure}
\centering
\includegraphics[width=\columnwidth]{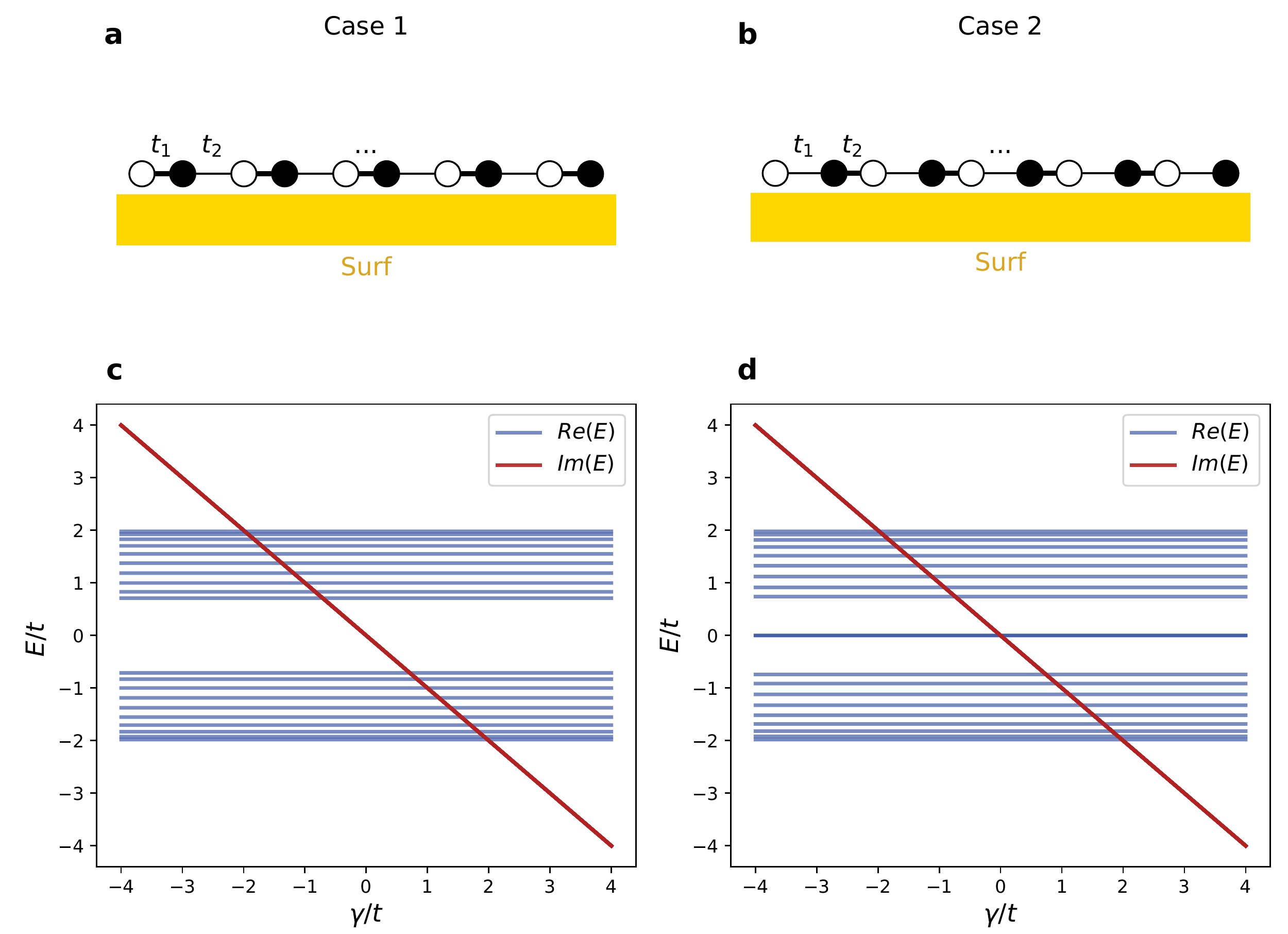}
\caption{Schematic representation of the SSH model coupled to a surface (in yellow) for (a) Case $1$ and (b) Case $2$. Their respective energy eigenvalues are shown in panels (c) and (d). Coupling to the surface is modeled by introducing the on-site energy ($-i \gamma$) to the SSH chain with the open boundary condition ($N=10$, 20 sites), for the same hopping parameters as \Figref{fig:fig1}.}
\label{fig:f2}
\end{figure}

For simplicity, we consider that the substrate induces an energy-independent imaginary self-energy on each site, characterized by the strength $\gamma$, and write
\begin{equation}
\Sigma = -i\gamma \sum_{m,n \in \{A,B\}} |m,n\rangle\langle m,n|.
\end{equation}
This corresponds to adding a uniform imaginary on-site potential $i\gamma$ across all sites, where $\gamma \neq 0$ represents the strength of the local decay ($\gamma > 0$) or gain ($\gamma < 0$) rate of particles (electrons) due to the existence of the substrate. The resulting effective Hamiltonian is
\begin{equation}
H_{\text{eff}} = H_{\text{SSH}} - i\gamma \sum_{m,n \in \{A,B\}} |m,n\rangle\langle m,n|.
\end{equation}
Since $H_{\text{eff}}$ is non-Hermitian, its eigenvalues $E_m$ generally become complex
\begin{equation}
H_{\text{eff}}|\psi_m\rangle = E_m |\psi_m\rangle, \quad E_m \in \mathbb{C}.
\end{equation}
After including the non-Hermitian self-energy terms, in the momentum space we have
\begin{equation}
H_{\text{eff}}(k) = \begin{pmatrix}
-i\gamma & t_1 + t_2 e^{-ik} \\
t_1 + t_2 e^{ik} & -i\gamma
\end{pmatrix}.
\end{equation}

From here, the complex band structure can be obtained by solving:
\begin{equation}
\det (H_{\text{eff}}(k) - EI ) = 0,
\end{equation}
to yield
\begin{equation}\label{eqn:compSSH}
E_{\pm}(k) = -i\gamma \pm \sqrt{t_{1}^{2} + t_{2}^{2} + 2 t_{1} t_{2} \cos(k)}.
\end{equation}

This equation shows that incorporating a uniform imaginary self-energy term $-i\gamma$ results in a global vertical shift of the energy spectrum in the complex plane. Explicitly, each eigenvalue $E_{\pm}(k)$ acquires a constant imaginary component $-i\gamma$, thereby translating the entire band structure upward or downward in the complex plane (depending on the sign of $\gamma$) along the imaginary axis without altering its dispersion in the real part of the spectrum. 

\Figref{fig:f2}(a,b) presents the schematics of this situation while the evolution of the energy eigenvalues as a function of $\gamma / t$ for a finite SSH chain with $N=10$ unit cells are shown in \Figref{fig:f2}(c, d), corresponding to Cases $1$ and $2$, respectively.
The two zero-modes associated with Case 2 associated with the terminal states are seen to remain centered at zero real energy.

\section{Partial decoupling from a substrate}

\begin{figure*}[!ht]
\centering
\includegraphics[width=\textwidth]{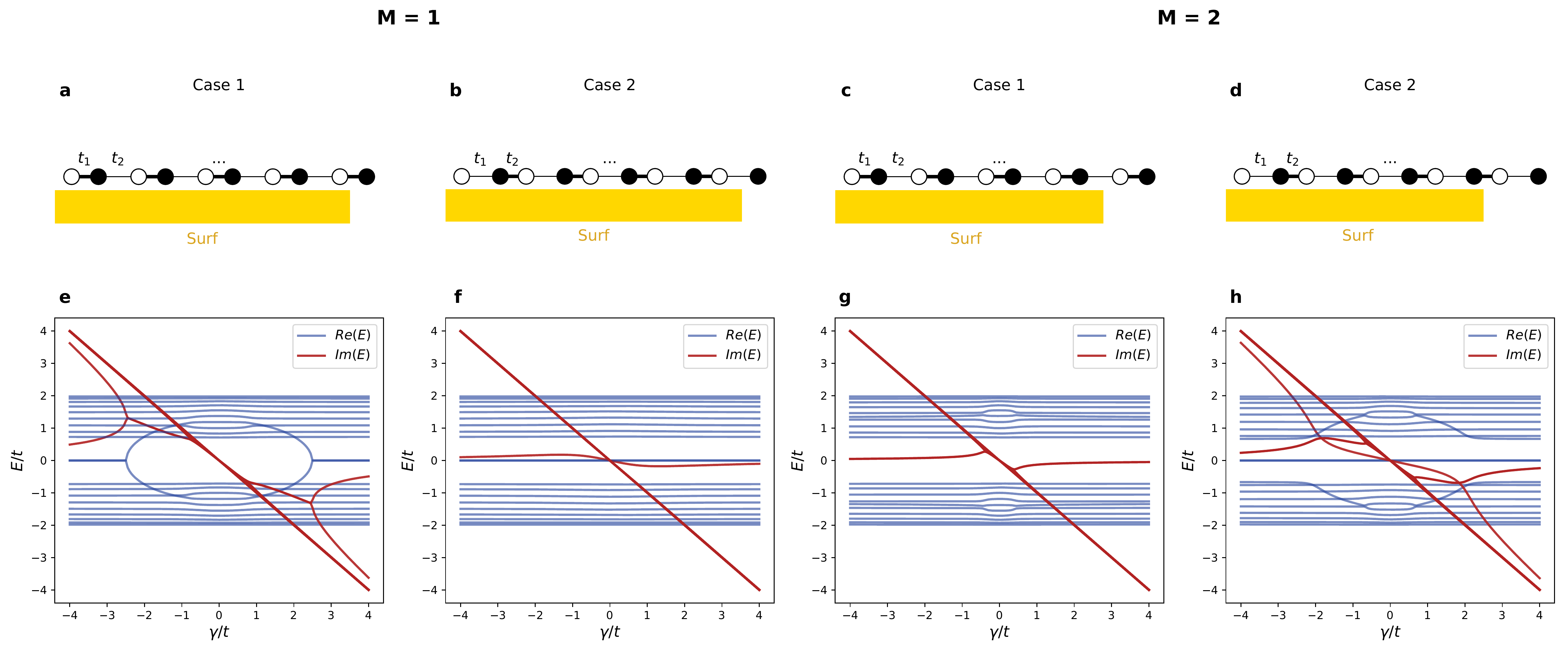}
\caption{(a,b) Schematics for the SSH chain with one decoupled site ($M=1$) for Case $1$ and Case $2$, respectively, and (c,d) similarly with two decoupled sites ($M=2$). The corresponding energy eigenvalues of their Hamiltonians for $N=10$ (20 sites) are shown in panels (e-h), respectively. The hopping parameters are as in \Figref{fig:fig1}. Decoupling the last site (e, f) results in the emergence of an exceptional point at $|\gamma| = 2|t_1|$ in Case $1$. Beyond the EP, zero-energy states appear in the real part of the energy spectrum, which decay ($\gamma > 0$) or gain ($\gamma < 0$ at two different rates. Decoupling two sites (g, h) results in the emergence of an exceptional point at $|\gamma| = 2|t|$ in Case 2.}
\label{f3}
\end{figure*}

Theoretically, one can imagine selective decoupling of molecular wires from a substrate, as an idealized modification of their interaction with the underlying surface. Inspired by experimental techniques involving scanning tunneling microscopes (STM) \cite{NIKLAS, eigler1990, hirjibehedin2006, Friedrich2024, Reecht2015, LaAmYu.09.ConductanceofSingle, jiang2022}, we consider a modified NH-SSH chain in which a specific section is treated as decoupled from the substrate. This decoupling is modeled by setting the corresponding coupling parameter $\gamma$ to zero. This idealized scenario provides a clear starting point to understand how NH effects and selective decoupling influence the electronic states.

\subsection{Disconnecting a few sites}

Decoupling the terminal atom of the SSH chain with the open boundary condition -- presented in \Figref{f3} (a, b) -- results in a modified Hamiltonian of the finite SSH chain in the form below
\begin{equation}
H_{\text{eff}} = 
\begin{pmatrix}
-i\gamma   & t_1        & 0          & \cdots & 0       & 0       \\[6pt]
t_1         & -i\gamma  & t_2        & \cdots & 0       & 0       \\[6pt]
0           & t_2        & -i\gamma  & \cdots & 0       & 0       \\[6pt]
\vdots      & \vdots     & \vdots     & \ddots & \vdots  & \vdots  \\[6pt]
0           & 0          & 0          & \cdots & -i\gamma & t_1     \\[6pt]
0           & 0          & 0          & \cdots & t_1     & 0
\end{pmatrix}.
\end{equation}
Here, all sites except the last one contain the $ -i\gamma_1 $ term, reflecting their coupling to the substrate (first environment). The last site, with $\gamma = 0$, represents a fully decoupled end of the chain. 

By gradually increasing the gain/loss parameter $\gamma$ relative to the average hopping amplitude $t$, we track the evolution of the energy eigenvalues and identify the conditions under which exceptional points emerge. For Case $1$, when $|\gamma| < 2.5|t|$, the system features two distinct states with different real parts but identical imaginary components. At $|\gamma| = 2.5|t|$, the real part of the two states merge at zero-energy part of the spectrum, accompanied by a bifurcation in the imaginary component, marking the occurrence of an EP. Increasing $|\gamma|$ beyond the EP leads to either decay ($\gamma > 0$) or growth ($\gamma < 0$) of these zero-energy states at two distinct rates as presented in \Figref{f3}(e).

In contrast, for Case $2$, decoupling the terminal atomic site does not create additional zero-energy states, see \Figref{f3}(f). Instead, a small imaginary component emerges in one of the zero-energy modes due to the site’s isolation from the lossy environment. This breaks the degeneracy and disrupts the chiral symmetry that characterizes the Hermitian nontrivial phase.

As shown in the Appendix \ref{appendix:2site}, a minimal two-site model with one atom connected to the substrate and the other decoupled from it, has a single EP at 
$\gamma = 2\,\lvert t^{\prime}\rvert$. By contrast, in our extended NH-SSH chain with partial decoupling, we numerically find an EP near $\gamma \approx 2.5\,t$, 
highlighting how boundary conditions, sublattice structure, and partial decoupling shift the degeneracy away from the two-site value.

\begin{figure*}[!ht]
\centering
\includegraphics[width=\textwidth]{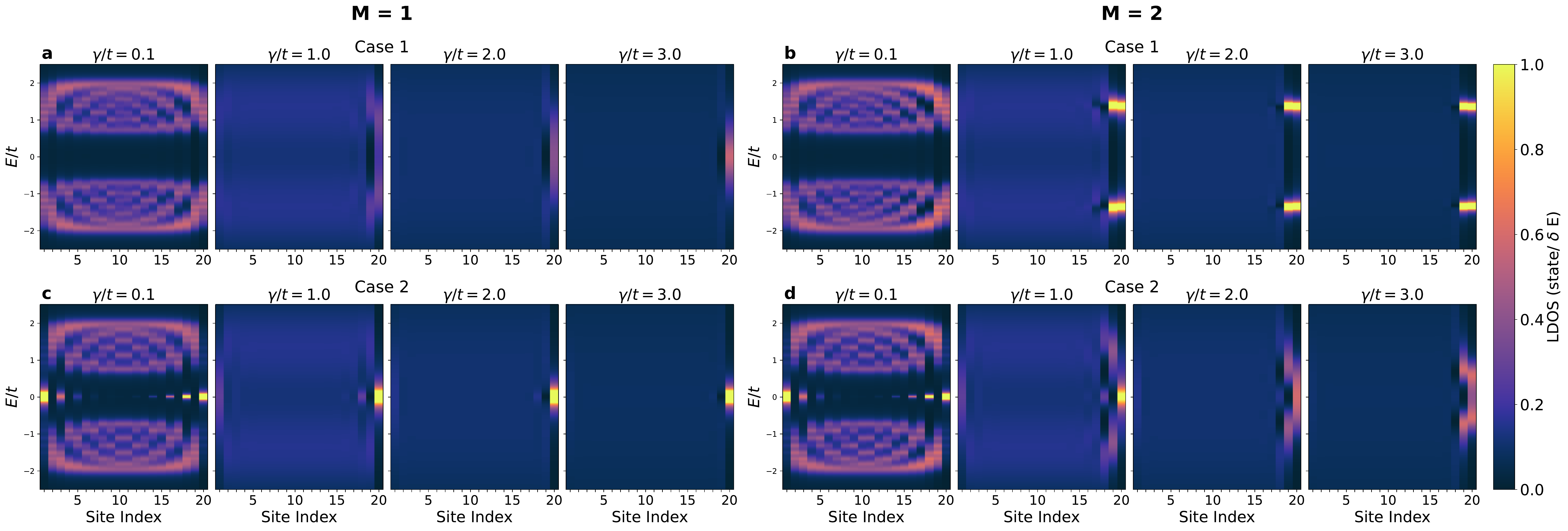}
\caption{LDOS maps for a lossy non-Hermitian SSH chain with length of $N=10$ (20 sites) and $M$ decoupled sites with hopping parameters as in \Figref{fig:fig1}. The top row (a, b) corresponds to Case $1$, and the bottom row (c,d) corresponds to Case $2$. Each column displays results for a different coupling strength $ \gamma / t = 0.1, 1.0, 2.0 $ and $3.0$. The vertical axis is the energy $E$, the horizontal axis the site index $m$, and the color scale indicates the LDOS magnitude.}
\label{fig:f4}
\end{figure*}

\begin{figure*}[!ht]
\centering
\includegraphics[width=\textwidth]{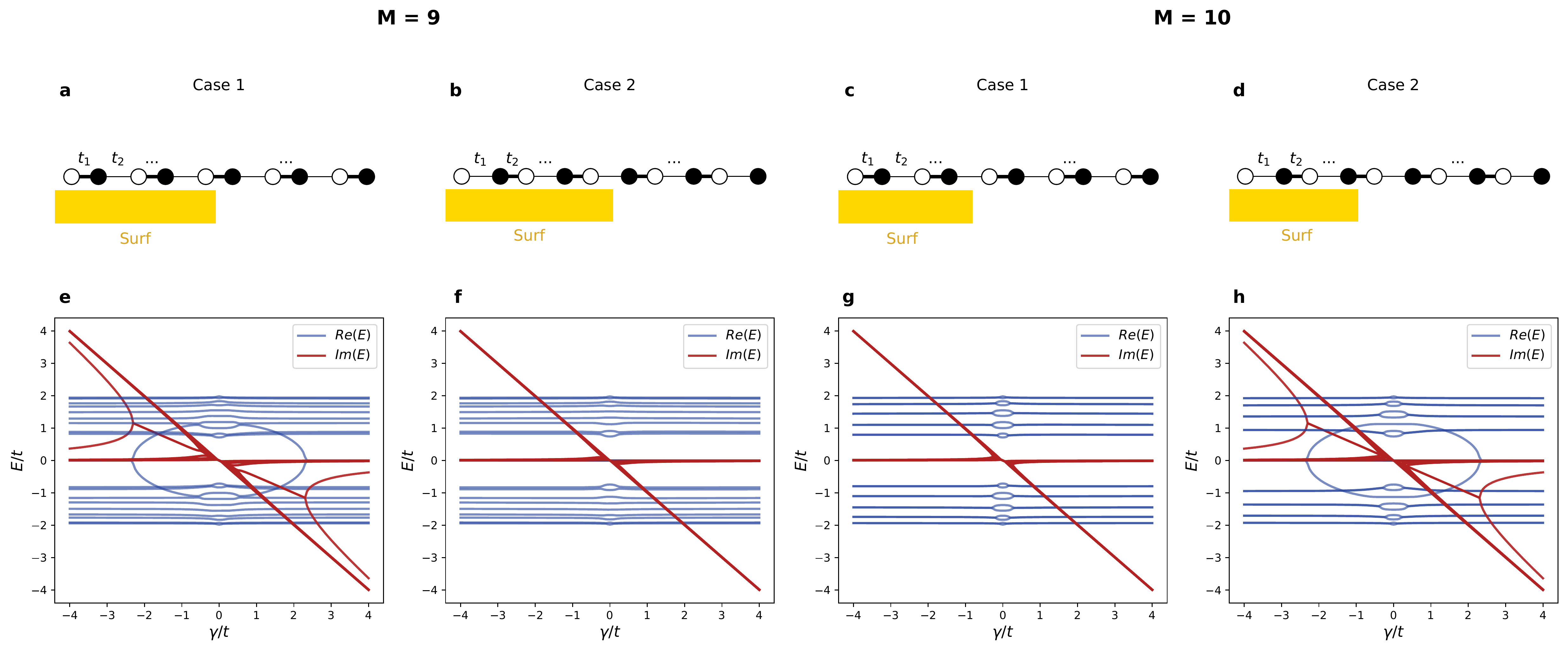}
\caption{Similar to \Figref{f3} but for $M=9$ and $M=10$, respectively.}
\label{fig:halfDecStates}
\end{figure*}

\begin{figure*}[!ht]
\centering
\includegraphics[width=\textwidth]{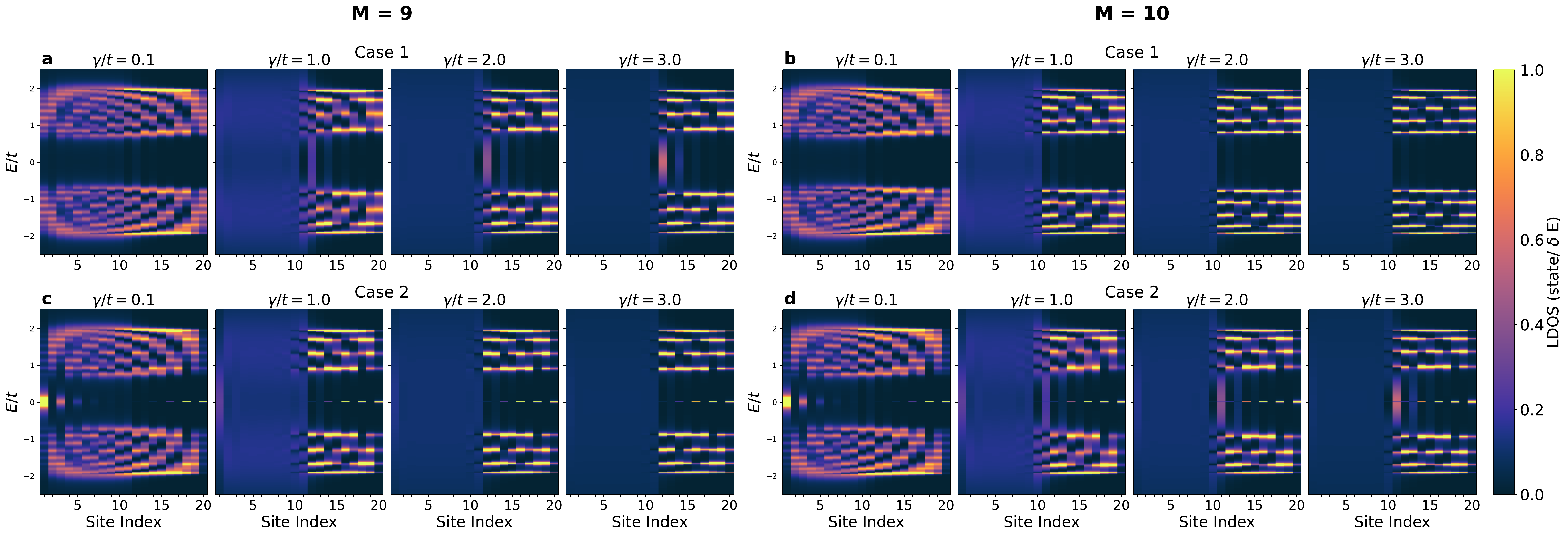}
\caption{Similar to \Figref{fig:f4} but for $M=9$ and $M=10$, respectively.}
\label{fig:ldosMid}
\end{figure*}

To obtain information and about the real space localization of the newly emerged states at zero energy in Case $1$, once again we employ the retarded Green’s function in the form of \Eqref{eqn:gFunction}. Subsequently, from the Green’s function we obtain the corresponding local density of states (LDOS) at site $m$ via
\begin{equation}
\rho_m(E) = -\frac{1}{\pi} \operatorname{Im}\bigl[G_{mm}(E)\bigr].
\end{equation}
In \Figref{fig:f4}, we present the result of these calculations as two-dimensional (2D) LDOS maps as a function of site index and energy for selected values of $\gamma$. In these plots, we focus on a lossy system with $\gamma > 0$. The local density of states, $\rho_m(E)$, is represented by a color scale to provide a more intuitive depiction of the spatial and energetic distribution of states.

Starting with Case $1$, in the weak coupling regime ($|\gamma| < 2.5 |t|$) we observe that at lower coupling values (e.g., $\gamma/t = 0.1$) the LDOS is (uniformly) distributed across multiple sites and energies, though compared to the freestanding chains, the energies now appear broadened. As $|\gamma|$ grows, the states on the last site move closer to the Fermi energy, and merge at the EP, see \Figref{fig:f4}(a). In the strong coupling regime ($|\gamma| > 2.5 |t|$), the states coupled to the surface essentially disappear and localize near the boundary at zero energy and the boundary state becomes almost delta-function-like. This can be traced back to the imaginary eigenvalues in \Figref{f3}(e) approaching the real axis as $|\gamma|$ grows \cite{ernzerhof}.

Analyzing Case $2$, we need to bear in mind that we have started with an SSH configuration which would normally exhibit zero-energy edge states localized at both ends of the chain. However, upon coupling the chain to a surface and increasing the coupling strength $(|\gamma|)$, we observe the zero-energy modes that were once robustly localized at both ends become asymmetrically pinned to a single edge, see \Figref{fig:f4}(c). This illustrates how coupling to an environment or surface -- by including NH effects -- can break the symmetric localization expected from a purely topological perspective. In other words, non-Hermiticity causes the topologically protected edge states to deform, damping or eliminating the topological state coupled to the substrate. Over time, the surface-bound edge state on the left vanishes, resulting in a monomode localized on the right-hand side of the chain. The decay rate of the bulk and the corresponding edge states can be further tuned by inserting fewer lossy sites, and more strategic placements \cite{slootman2024}.

\subsection{Disconnecting half of the chain}

By progressively decoupling additional lattice sites from the end of the chain (right-hand side) and modifying their corresponding on-site imaginary energies in the effective Hamiltonian from $-i\gamma$ to zero, we uncover a distinct pattern in the formation of boundary states within the two regimes of the non-Hermitian SSH model. Analysis of the electronic states (\Figref{fig:halfDecStates}) and local density of states (\Figref{fig:ldosMid}) reveals that the emergence of zero-energy boundary states is strongly dependent on the number of decoupled sites.

 In Case $1$, decoupling an odd number of sites results in the formation of a zero-energy peak at the EP, with the corresponding boundary state localized at the edge of the chain, see \Figref{fig:ldosMid}(a). This zero-energy feature is born at the EP and becomes increasingly pronounced under strong coupling, as $\lvert\gamma \rvert$ exceeds $2|t_1|$. However, when an even number of sites is decoupled, the boundary state vanishes as shown in \Figref{fig:ldosMid}(b). 

In contrast, for Case $2$, decoupling an even number of sites induces the formation of a zero-energy boundary state at the EP, see \Figref{fig:ldosMid}(d). Under strong coupling, this zero-energy peak persists and is accompanied by a previously reported monomode localized at the edge of the decoupled region. In this regime, decoupling an odd number of sites suppresses the boundary state (\Figref{fig:ldosMid}(c)). Notably, the zero-energy mode associated with the suspended part of the chain remains intact regardless of the number of decoupled sites.

\section{Transmission to a tip electrode}
\begin{figure*}[!ht]
\centering
\includegraphics[width=\textwidth]{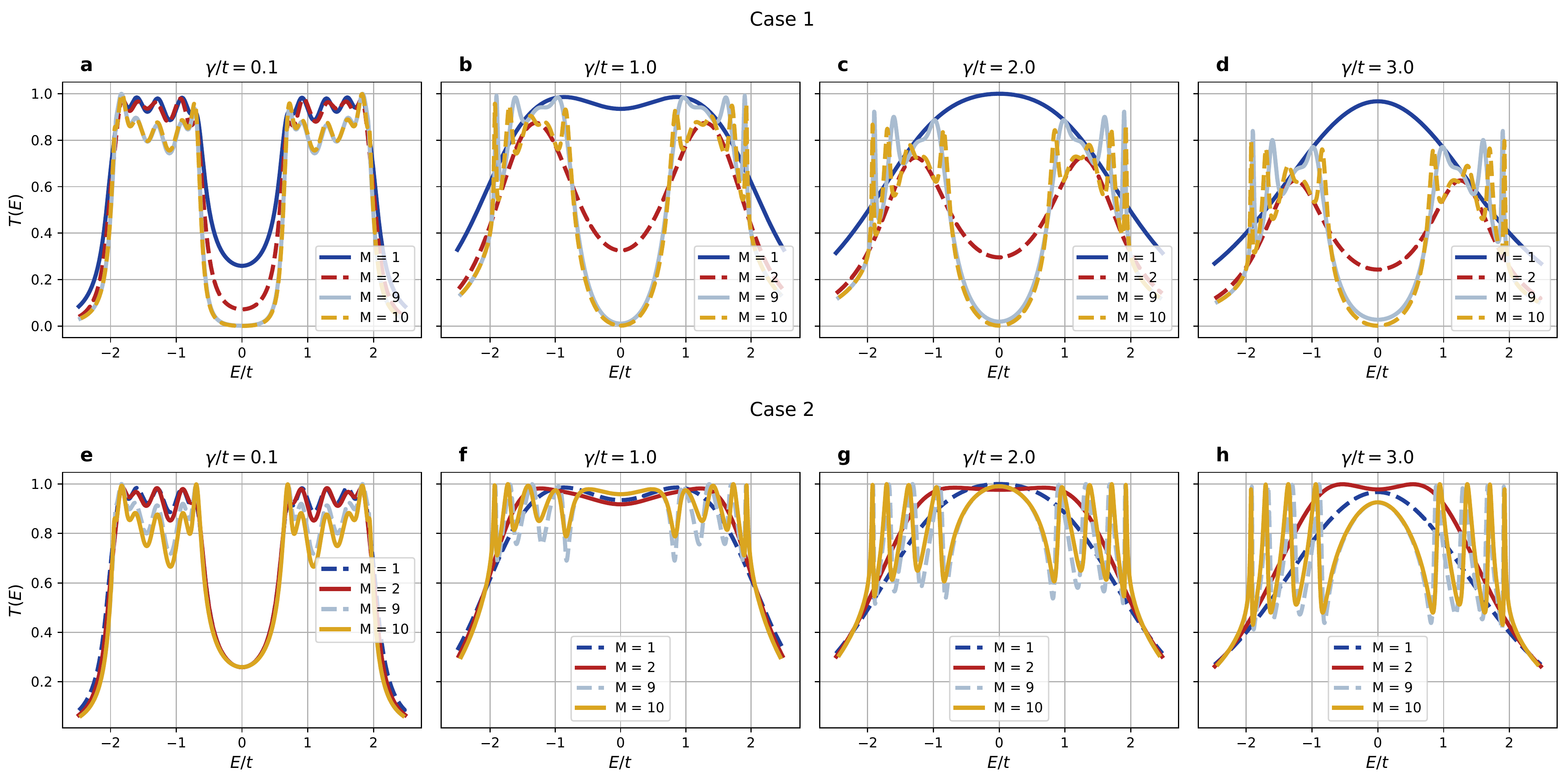}
\caption{Electron transmission $T(E)$ across the SSH chain from substrate to an electrode attached to the terminal site for Case $1$ (top row) and Case $2$ (bottom row), respectively. In Case $1$, the finite SSH chain consists of $N=10$ unit cells with parameters $t_1=4/3$, $t_2=2/3$, and $\gamma_2=1$. In Case $2$, the chain also consists of $N=10$ unit cells but with $t_1=2/3$, $t_2=4/3$, and $\gamma^{\prime}/t=1$. Panels $(a-d)$ and $(e-h)$ depict the results for $\gamma=0.1$, $\gamma=1.0$, $\gamma=2.0$, and $\gamma=3.0$, respectively. Different colors in each panel represent varying numbers of decoupled sites $M$ at the ends of the chain. For Case $1$, in the strong coupling regime, zero-energy peaks emerge only for odd $M$, while they are absent for even $M$. In Case $2$, strong coupling results in a zero-energy monomode and weaker transmission for odd $M$, whereas even $M$ enhance transmission due to the presence of a boundary state.}
\label{fig:transmish}
\end{figure*}

To analyze the electronic transport of the partially decoupled NH-SSH chain, we employ the Landauer-Büttiker formalism. In this approach, the central region (chain) is connected to two leads: a substrate (lead $1$) and a tip (lead $2$). The transport properties are captured by the energy-dependent transmission function $T(E)$, which quantifies the probability that an electron with energy $E$ injected from one lead will be transmitted to the other. The energy-dependent transmission function for this scenario is given by
\begin{equation}
T(E) = \mathrm{Tr}[\Gamma_\text{sub} \, G(E) \, \Gamma_\text{last}  \, G^\dagger(E)],
\end{equation}
where $G(E)$ is the Green's function of the central region, and $\Gamma_\text{sub} $, $\Gamma_\text{last} $ are the coupling (broadening) matrices for the surface and last-atom electrodes, respectively.
For simplicity, we take into account energy-independent imaginary self-energies, so the transmission function becomes
The second electrode is attached only to the last atom, with strength equal to $\gamma^{\prime}=t$, so we have
\begin{equation}
\Gamma_\text{last} \;=\; 
\begin{pmatrix}
0 & & & \\
  & \ddots & & \\
  & & 0 & \\
  & & & 2 \gamma^{\prime}
\end{pmatrix}.
\end{equation}

We present the transmission results as a function of energy in \Figref{fig:transmish}. The finite SSH chain used in these calculations also has 
$N=10$ unit cells. In Case~1, upon coupling the chain to the tip ($M = 1$), we observe 
two peaks for $\lvert\gamma\rvert < 2\lvert t\rvert$ which merge at $\lvert\gamma\rvert = 2\lvert t\rvert$. Increasing the coupling strength leads to the 
emergence of a sharper zero-energy peak in $T(E)$, along with a slight dip in the transmission probability. We demonstrate that partially decoupling more atoms of the Case~1 chain leads to the absence (for even $M$) and emergence (for odd $M$) of zero-energy electron transmission. However, even if an interface state is expected for $M=9$ under Case~1, its influence on the transmission remains negligible. This arises 
from the fact that the second electrode is coupled to the far end of the chain, where that localized state does not extend. The bottom row of \Figref{fig:transmish} contains the transmission results for Case~2. Conversely, in this case electron transmission at zero energy is increased (decreased) at the $2|t|$ and beyond for even (odd) number of decoupled sites $M$.

\section{Conclusions}

In this work, we have demonstrated how coupling a one-dimensional Su–-Schrieffer-–Heeger chain to a substrate induces a controlled non-Hermiticity that profoundly alters its electronic structure and topological features. By systematically varying the substrate coupling strength and selectively decoupling sites at one end of the chain (mimicking an STM tip lifting atoms or placing part of the chain on a second substrate), we revealed the emergence of robust zero-energy boundary modes and observed how they evolve and gain strength with respect to the exceptional points. 

In particular, we demonstrated that if the intracell hopping energies are larger than intercell hopping amplitudes ($|t_1| > |t_2|$), lifting an odd number of sites from the substrate can sustain sharply localized zero-energy modes, whereas even-numbered decoupling generally suppresses these resonances. In the opposite case, when the intercell hoppings are larger than intracell hopping amplitudes ($|t_1| < |t_2|$), an odd number of decoupled sites damps the electronic states of the chain except for the suspended part, leading to monomodes with large amplitudes at zero energy. In this case, an even number of decoupled sites gives rise to zero-energy states localized at the boundary in addition to the monomode, as observed via real-space LDOS maps.

Furthermore, our analysis of the transmission function indicates that these decoupled states can serve as conduits for near-zero-energy transport, provided that the parity (odd vs.\ even) of the lifted sites maintains the boundary mode. Beyond extending the usual SSH model into a non-Hermitian regime, our results highlight that one can selectively tune resonant edge states by engineering both the coupling strength and the spatial extent of the decoupled segment. This interplay of non-Hermitian effects, exceptional points, and boundary-mode manipulation opens new avenues for designing quantum devices and sensors based on robust edge states. 

Future studies may exploit these findings in more realistic multi-band systems or explore dynamic control of substrate coupling to achieve reconfigurable, non-Hermitian topological phases in on-surface–synthesized nanomaterials. For instance, rather than assuming a uniform imaginary potential $i\gamma$ across all adsorbate sites, one can define $i\gamma(m)$ that changes from site to site, capturing effects of local adsorption geometry, defects, or impurities. This leads to a spatially inhomogeneous decay/gain profile, making some sites more weakly coupled and others more strongly coupled. Consequently, boundary modes and near‐zero‐energy resonances can shift or broaden in the complex energy plane, and exceptional points may emerge or disappear at different parameter values compared to the uniform case.

\section{Acknowledgements}
We express our gratitude to Dr. Dario Bercioux, Dr. Niklas Friedrich, Dr. Geza Giedke, Dr. Nico Leumer, and Carolina Martinez Strasser for their insightful suggestions and valuable discussions during the course of this work.
This research was supported by the IKUR Strategy under the collaboration agreement between the Ikerbasque Foundation and DIPC on behalf of the Department of Education of the Basque Government, the Spanish MCIN / AEI / 10.13039/501100011033 (PID2023-146694NB-I00, GRAFIQ), and the Basque Department of Education (PIBA-2023-1-0021, TENINT).
\vspace{1cm}

\appendix*

\section{Minimal 2-Site Example for an Exceptional Point}
\label{appendix:2site}

We consider a 2-site Hamiltonian
\begin{equation}
H \;=\; 
\begin{pmatrix}
-\,i\gamma & t^{\prime} \\
t^{\prime}         & 0
\end{pmatrix},
\end{equation}
for which only the first site has a NH part set by $\gamma$.
We seek the value(s) of $\gamma$ at which an EP occurs, i.e., where the matrix has a repeated eigenvalue and coalescing eigenvectors.
The characteristic polynomial is given by
\begin{equation}
p(E) \;=\;
\det(H - E\,I) \;=\;
\begin{vmatrix}
-\,i\gamma - E & \; t^{\prime} \\
t^{\prime}              & \; -\,E
\end{vmatrix}.
\end{equation}
Which is:
\begin{equation}
p(E) \;=\;
E^2 + i\gamma\,E - t^{\prime2}.
\end{equation}
%
%
A repeated (double) root of the polynomial $p(E)$ occurs when the discriminant is zero. For a quadratic
\begin{equation}
E^2 + bE + c,
\end{equation}
the discriminant is $\Delta = b^2 - 4c$. Here $b = i\gamma$ and $c = -\,t^{\prime2}$, so
\begin{equation}
\Delta \;=\;
-\,\gamma^2 \;+\; 4\,t^2.
\end{equation}
Thus, the EP occurs at
\begin{equation}
\gamma = 2\,\lvert t\rvert.
\end{equation}

\bibliography{references.bib}

\end{document}